\documentclass[aps,prd,preprint,showpacs,floatfix,preprintnumbers,nofootinbib]{revtex4}
\usepackage{pifont}
\usepackage{float}
\usepackage{placeins}
\usepackage[utf8]{inputenc}
\usepackage[vcentermath,enableskew]{youngtab}
\usepackage{fancybox}
\usepackage{hhline}
\usepackage{dcolumn}
\usepackage{textcomp}
\usepackage{epsfig,graphics,graphicx}
\usepackage{amsfonts,amssymb,amsmath}
\usepackage{pifont}
\usepackage{bm}
\usepackage{longtable}
\usepackage{appendix}
\usepackage{lscape}
\usepackage[mathscr]{euscript}
\usepackage{mathrsfs}
\usepackage{multirow}
\usepackage{rotating}
\usepackage{color}   
\usepackage[dvipsnames]{xcolor}

\newcommand{\fr}{\frac}
\newcommand{\beq}{\begin{equation}}
\newcommand{\eeq}{\end{equation}}
\newcommand{\bea}{\begin{eqnarray}}
\newcommand{\eea}{\end{eqnarray}}

\newcommand{\ord}[1]{{\cal{O}}\left( #1 \right)}

\newcommand{  \xmark }{\ding{55}}%

\DeclareFontFamily{OT1}{pzc}{}
\DeclareFontShape{OT1}{pzc}{m}{it}%
              {<-> s * [0.900] pzcmi7t}{}
\DeclareMathAlphabet{\mathpzc}{OT1}{pzc}%
                                 {m}{it}

\DeclareMathAlphabet{\mathcalligra}{T1}{calligra}{m}{n}

\usepackage{hyperref}
\hypersetup{pdfauthor={me}, colorlinks=true, citecolor=blue, urlcolor=blue, linkcolor=black}

\begin{document}

\preprint{\vbox{\hbox{ JLAB-THY-12-1644} }}

\vspace*{2cm}

\title{
Global analysis of the  negative parity non-strange baryons  in the $1/N_c$ expansion
}
\author{
E. Gonzalez de Urreta$^{a,b}$,
J. L. Goity$^{c,d}$,
N. N. Scoccola$^{a,b,e}$}

\address{
$^a$
Department of Theoretical Physics,
Comisi\'on Nacional de Energ\'{\i}a At\'omica, 1429  Buenos Aires, Argentina.\\
$^b$CONICET, Rivadavia 1917, (1033) Buenos Aires, Argentina.\\
$^c$Department of Physics, Hampton University, Hampton, VA 23668, USA.\\
$^d$Thomas Jefferson National Accelerator Facility, Newport News, VA 23606, USA.\\
$^e$Universidad Favaloro, Sol{\'\i}s 453,  1078 Buenos Aires,
Argentina. }


\begin{abstract}
A global study of the negative parity non-strange baryon observables is performed in the framework of the $1/N_c$ expansion. Masses, partial decay widths and photo-couplings  are simultaneously analyzed. A main objective is to determine the composition of the spin $\frac{1}{2}$ and $\frac{3}{2}$ nucleon states, which come in pairs and involve two mixing angles which can be determined and tested for consistency by the mentioned observables. The issue of the assignment of those nucleon states to the broken $SU(4)\times O(3)$ mixed-symmetry multiplet is studied in detail, with the conclusion that the assignment made in the old studies based on the non-relativistic quark model is the preferred one. In addition, the analysis involves an update of the input data with respect to previous works.
\end{abstract}

\maketitle

\section{Introduction}

The excited baryon states that correspond in the constituent quark
model to the first radial and orbital excitations fit  well
into respectively a positive parity symmetric  and a
negative parity mixed-symmetric irreducible representation of
the spin-flavor group $SU(2 N_f)$, where $N_f$ is the number of light flavors. For $N_f=2$  both representations are 20-dimensional
and are respectively denoted by $\mathbf{20}$ and  $\mathbf{20'}$.
The negative parity non-strange excited baryons states are empirically  the best known ones  \cite{Beringer:1900zz},
and are organized in the $[{\mathbf{20'}},\ell=1]$ representation of  $SU(4)\times O(3)$ \cite{close}.
Their masses, partial decay widths and photo-couplings are known to a degree where detailed analyses as the one presented here can be carried out and lead to some clear conclusions.
  The states consist of two spin 1/2 nucleon states: N(1535) and N(1650), two spin 3/2 nucleon states: N(1520) and N(1700), one spin 5/2 nucleon state:
  N(1675) and two isospin multiplets: $\Delta(1620)$ and $\Delta(1700)$ of spin 1/2 and 3/2 respectively.
   A very important issue, central to the study in this work, is  the assignment of the physical states to the pairs of  spin 1/2 and 3/2 nucleon states, and also the accurate determination of the corresponding mixing angles $\theta_{2J=1}$ and $\theta_3$.  In order to settle that issue in as model independent as possible fashion it  is necessary to carry out a simultaneous analysis of masses, partial decays widths  and photo-couplings, as presented in this work. The framework used here is the $1/N_c$ expansion as presented elsewhere for masses \cite{Carlson:1998gw,Carlson:1998vx}, decays \cite{Goity:2004ss} and photo-couplings \cite{Scoccola:2007sn}.

  The advantage of using the $1/N_c$ expansion \cite{tHo74}
   is that, in principle, the analysis is based on a framework which is true to QCD. It provides a justification for the approximate spin-flavor symmetry observed in baryons \cite{Gervais:1983wq,Gervais:1984rc,Dashen:1993as,Dashen:1993ac},  which permits the implementation of the $1/N_c$ starting from that symmetry limit.
   From multiple applications it transpires that the $1/N_c$ expansion works well in the real world with $N_c=3$ \cite{Lucini:2012gg}.  In particular,  its implementation in the analysis of excited baryons (see e.g. Refs. \cite{Carlson:1998gw,Carlson:1998vx,Goity:2004ss,Scoccola:2007sn,car94,car98,Schat:2001xr,Cohen:2003tb,mat05,Goity:2003ab}), has shown to be phenomenologically  consistent in the sense that  the sub-leading corrections are of natural size.
In the present work, the  $1/N_c$ expansion is implemented as an operator expansion, where the coefficients are determined by the QCD dynamics, and those coefficients are evaluated by fitting to the pertinent observables. In addition, since usually at the first sub-leading order the number of observable quantities exceeds the number of operators, there are relations  independent of the coefficients, which are exact up to sub-sub-leading order corrections, and which should serve as a good test of the  framework.  The physics of excited baryons thus gets sorted out  hierarchically in powers of $1/N_c$, helping in this way to define and organize the discussion of the different effects.
   It should be noted that the implementation for the negative parity baryons is carried out under the assumption that $SU(4)\times O(3)$ configuration mixings, i.e. mixings of different irreducible representations of that group, can be disregarded. A discussion of such effects was presented in \cite{Goity:2004pw}. In the $1/N_c$ expansion, baryon observables are represented by effective operators which are themselves decomposed in an operator basis which is organized according to the $1/N_c$ power counting of the operators. The coefficients of that decomposition are determined by the QCD dynamics. These are obtained by fitting the corresponding observables to experimental data.


  In the case of the negative parity baryons, the analysis of the masses   suffers from the four-fold ambiguity mentioned above in the assignment of the spin 1/2 and 3/2 nucleon physical states to the theoretical ones. It turns out that such ambiguity cannot be resolved by analyzing the masses alone, as acceptable fits, i.e. $\chi^2_{dof}\sim 1$ can be obtained for all four assignments.
   Because the strong decay partial widths with emission of a single pion, and the EM transitions are more sensitive to those assignments, they must be used to arrive at a definite conclusion. Historically, the first such analysis was carried out in the framework of the quark model \cite{Koniuk:1979vy}. In the $1/N_c$ expansion the work of Pirjol and Schat \cite{Pirjol:2003ye} was the first to address the problem , and it was also piecemeal addressed in other works on strong decays \cite{Goity:2004ss} and EM transitions \cite{Scoccola:2007sn}. However, an exhaustive combined analysis was still lacking, which is now presented in this work.


The present work is organized as follows. Sec.\ II  provides a
brief review of the framework used to analyze the non-strange
low-lying negative parity excited baryons in the framework of
the $1/N_c$ expansion of QCD. Some parameter independent
relations valid to ${\cal O} (N_c^0)$ are also presented and
discussed in this section. Sec. III describes the procedure
followed to perform the global fits of the empirically
known baryon properties to ${\cal O} (1/N_c)$ and analyze the
corresponding results.
The main conclusions are presented in Sec. IV. Finally, includes two Appendices:  Appendix  I
gives the lists of operators furnishing the bases for the analysis of the baryon
properties considered in this work, while Appendix II gives some parameter independent relations
satisfied at LO  by the photo-couplings.

\section{Properties of the low-lying non-strange negative parity baryons in the $1/N_c$ expansion}

In the following a brief description of the    $1/N_c$ expansion   excited baryon  framework is presented.

The  group $SU(4)$  has 15 generators, namely, the spin, isospin and  spin-isospin generators:
{$S^i$,$T^a$,$G^{ia}$} , with $i=1,2,3$ and $a=1,2,3$, normalized to fulfill the following commutation relations:
\begin{eqnarray}
\left[S^i,S^j\right]&=&i\epsilon^{ijk}S^k,
\left[T^a,T^b\right]=i\epsilon^{abc}T^c,
\left[S^i,T^a\right]=0, \nonumber\\
\left[G^{ia},G^{jb}\right]&=&\frac{i}{4}\delta^{ij}\epsilon^{abc}T^c+\frac{i}{4}\delta^{ab}\epsilon^{ijk}S^k,\\
\left[S^i,G^{ja}\right]&=&i\epsilon^{ijk}G^{ka},
\left[T^a,G^{ib}\right]=i\epsilon^{abc}G^{ic}. \nonumber
\end{eqnarray}
The matrix elements of $G^{ia}$  between baryons with spin of order $N_c^0$ are $\ord{N_c}$.

Introducing the spherical notation for the group generators and omitting the projections for simplicity, the generators are denoted by:
\begin{equation}
S^{[1,0]},\qquad T^{[0,1]}\qquad {\rm and} \qquad G^{[1,1]}~,
\end{equation}
where the notation $\Lambda^{[J,I]}$ implies that the operator
$\Lambda$ transforms as the $[J,I]$ irreducible representation of
the $SU(2)\times SU(2)$ spin$\times$isospin subgroup of $SU(4)$.

\subsection{$\mathbf{[20', 1^-]}$ states for arbitrary $N_c$}


For arbitrary $N_c$, the states of interest here belong to the SU(4) mixed-symmetric representation $(p,q,r)=(N_c-2,1,0)$, which   has dimension $((N_c-1)(N_c+2)!)/2 N_c!)$.
These states are constructed  by coupling a fundamental $SU(4)$ state to a totally symmetric state with $N_c-1$ indices, which is denoted as the {\it core}  \cite{Carlson:1998vx}:
\begin{equation}
\overbrace{\yng(2)\ldots\yng(1)}^{N_c-1}\otimes\,\young(~)=
\overbrace{\stackrel{\bf S}{\yng(2)\ \ \ldots\ \ \young(~)}}^{N_c}
\oplus\overbrace{\stackrel{\bf MS}{\young(\,\,,~)\ \
\raisebox{5.5pt}{\ldots} \ \ \raisebox{2.5pt}{\yng(1)}}}^{N_c-1}\,\,\,,
\end{equation}
Denoting the core states by  $\mid S_c,S_{c3},I_{c3}\rangle$, where these states belong to the symmetric $SU(4)$ representation in which  $I_c=S_c$ is satisfied, the mixed-symmetric states of interest  denoted by $\mid S, S_3;I, I_3\rangle $ are  given by   \cite{Carlson:1998vx}:




\begin{eqnarray}
\mid S, S_3; I , I_3\rangle_{\mbox{\bf MS}} &=&\sum_{\xi=\pm \frac{1}{2}}
C_{\mbox{\bf MS}}(I,S,\xi) \mid S, S_3 ; I , I_3 ; S_c=S+\xi \rangle ,
\end{eqnarray}
where
\begin{eqnarray}
C_{\mbox{\bf MS}}(I,S, \pm \frac{1}{2})&=&\left\{
    \begin{array}{c}
      1~~{\rm si}~~I=S \pm 1   \\
      0~~{\rm si} ~~ I=S \mp 1 \\
      \pm \sqrt{\frac{(2 S + 1 \pm 1 ) (N_c+ 1 \pm (2 S+ 1))}{2 N_c ( 2 S + 1)} } ~~{\rm si} ~~I=S
    \end{array}\right.~,
\end{eqnarray}
and
\begin{eqnarray}
\mid S,  S_3 ; I , I_3 ; S_c\rangle &=&\sum_{\,s_3,i_3} \langle S_c, S_3-\,s_3; 1/2, \,s_3  \mid S, S_3\rangle
\langle I_c=S_c, I_3-i_3; 1/2, i_3  \mid I, I_3 \rangle \nonumber\\
&\times&
   \mid S_c ,  S_3-\,s_3 ; I_c=S_c, I_3-i_3 \rangle \,\otimes \mid 1/2, \,s_3 ; 1/2, i_3   \rangle \label{states}.
\end{eqnarray}
 $S$ is the total spin of the baryon associated with the spin group $SU(2)$, $S_c$ is the core's spin.
 The $SU(2)$ Clebsch-Gordan coefficients are defined with the standard Condon-Shortley phase convention (see e.g. Ref. \cite{edmonds}).

The states belonging to the $[\mathbf{20'},1^{-}]$ of $SU(4)\times O(3)$ are now expressed by including the orbital angular momentum,
\begin{equation}
\mid J , J_3 ; I , I_3 ; \ell , S \rangle_{\mbox{\bf MS}} =
\sum_m \langle   \ell, m ; S , J_3 - m \mid J, J_3\rangle \; \mid \ell , m \rangle
\,\otimes \mid S, J_3 - m ;I , I_3 \rangle_{\mbox{\bf MS}}.
\label{estados}
\end{equation}

The symmetric states are given explicitly in Ref. \cite{Goity:2004ss}.

The state mixings mentioned in the Introduction are  induced by interactions that break spin/isospin symmetry.
These mixings are between the pairs of states that are in the two octets with the same $J=\frac{1}{2}$ or $\frac{3}{2}$. The mixings are defined according to:

\begin{equation}
\left(\begin{array}{c} N_J \\ N'_J \end{array} \right) =
\left(\begin{array}{cc}  \cos\theta_{2J} & \sin\theta_{2J} \\
                       -\sin\theta_{2J} & \cos\theta_{2J}
\end{array}\right)
\left(\begin{array}{c} ^2N^*_{J} \\ ^4N^*_{J}\end{array} \right)\,\,,
\label{conv}
\end{equation}
where $N^{(')}_J$ are mass eigenstates. The mixing angles are determined by the diagonalization of the corresponding  $2\times 2$ mass matrices. The
eigenvalues and mixing angles are   given by:
\begin{equation}
m_{N_J ,N'_J }  = \frac{{M_2^J  + M_4^J }}{2} \pm {\rm sign}(M_2^J-M_4^J)\; \sqrt {\left( {\frac{{M_2^J  - M_4^J }}{2}} \right)^2
+ \left( {M_{24}^J } \right)^2 }\,\,,
\end{equation}
\begin{equation}
\theta_{2J}=\frac{1}{2}\,\arctan\left(\frac{2M^{J}_{24}}{M^{J}_{2}-M^{J}_{4}}\right) ({\rm mod} \;\pi)\,\,,
\end{equation}
where $M_2^J = \langle {}^2N_J^*| \textbf{M} |{}^2N_J^* \rangle$, etc.  Angles can be defined to be in the interval $[0,\pi)$ by conveniently choosing the phase of the physical  states.

As already mentioned there are four possible assignments of the states $N_J$ and $N^{'}_J$ to the physical ones. They are:
\begin{eqnarray}
\mbox{Set 1} & = & \left\{ N_{\frac12}=N(1535), ~N'_{\frac12}=N(1650), ~N_{\frac32}=N(1520),~N'_{\frac32}=N(1700) \right\}
\nonumber \\
\mbox{Set 2} & = & \left\{ N_{\frac12}=N(1650), ~N'_{\frac12}=N(1535), ~N_{\frac32}=N(1520),~N'_{\frac32}=N(1700) \right\}
\nonumber \\
\mbox{Set 3} & = & \left\{ N_{\frac12}=N(1535), ~N'_{\frac12}=N(1650), ~N_{\frac32}=N(1700),~N'_{\frac32}=N(1520) \right\}
\nonumber \\
\mbox{Set 4} & = & \left\{ N_{\frac12}=N(1650), ~N'_{\frac12}=N(1535), ~N_{\frac32}=N(1700),~N'_{\frac32}=N(1520) \right\}
\label{sets}
\end{eqnarray}

It should be noted that in previous works  \cite{Carlson:1998gw,Carlson:1998vx,Pirjol:2003ye}
the mixing angles have been defined using some specific state assignment. For example,
in Ref. \cite{Pirjol:2003ye} they are defined in their Eqs. (18) and (19),
which only correspond to the definition Eq. (10) when states are assigned according to Set 1 in Eq. (\ref{sets}).

\subsection{Operator analysis}

A local QCD color singlet operator, expressed in terms of the quark and gluon fields and  which has irreducible transformation properties under isospin and rotations, can be expanded in the subspace of baryon states in powers of $1/N_c$ as an operator expansion consisting of composite operators built with
products of  operators which are tensors of $SU(4)\times O(3)$ \cite{Dashen:1994qi,Goity:1996hk}. Those tensor operators have a well
defined $1/N_c$ power counting, as described below.
Thus, a QCD  operator has the $1/N_c$ expansion of the form:
\begin{equation}
\mathcal{O}_{QCD}=\sum_{n,\ell}C^{(n)}_{\ell}\mathcal{O}^{(n)}_{\ell}.
\label{eq:counting}
\end{equation}
The $1/N_c$ order of the composite operators is determined by the $n$-body character or degree of these, which is  given  by the number of quark fields necessary to built an  operator with the same transformation properties \cite{Manohar:2001cr}, and by the $N_c$ order of the matrix elements of the tensor operators appearing as factors. Since an $n$-body operator requires the exchange of $n-1$ gluons to be produced, there is an overall factor $1/N_c^{n-1}$, and thus the (naive) order in $1/N_c$ of an $n$-body  composite operator becomes:
\beq
\nu(\mathcal{O}^{(n)})=n-1-\kappa,
\label{kapa}
\eeq
where $\kappa$ is given  by the order of the factors, and $n$ satisfies $n\leq N_c$.
The coefficients $C^{(n)}_{\ell}$ in Eq. (\ref{eq:counting}) are the unknown dynamical coefficients
of $\ord{N_c^0}$, which are determined by the QCD dynamics.

In the subspace of the \textbf{MS} states of interest, a basis of composite operators can be constructed by  distinguishing two sets of generators according
to whether they act on the core or the ``excited quark"  which corresponds to the spin-flavor index associated to the second row of the Young tableaux.
Generators acting on the core will be denoted by $\left\{S_c^{[1,0]}, T_c^{[0,1]}, G_c^{[1,1]}\right\}$ and those
acting on the excited quark by $\left\{\ell^{[1,0]},s^{[1,0]},t^{[0,1]},g^{[1,1]}\right\}$.
Then, $n$-body operators can be those consisting of the product of  $n$ $SU(4)$ generators acting only  on the core,
or $n-1$ $SU(4)$ generators acting on the core and at least one generator of
$SU(4)\times O(3)$ acting on the excited quark. It should be noted that the order
$\kappa$ appearing in Eq. (\ref{kapa}) takes the value
$\kappa=0$ for all the generators except the generator $G_c$ for which $\kappa=1$.

\subsection{Mass operators}

The   basis of mass operators up to order $1/N_c$ is taken from that obtained in \cite{Carlson:1998vx}, and shown here in Table \ref{masop} of Appendix I.  Convenient normalization factors have been included in the definition of the basis operators in such a way that they have natural size matrix elements. There is one operator $\ord{N_c}$, namely the identity or baryon number. There are only two operators $\ord{N_c^0}$, both involving the coupling of the angular momentum $\ell$. The first corresponds in the quark model picture to the Thomas precession spin-orbit interaction, which for baryons in $\mathbf{S}$ representation is $\ord{1/N_c}$ but in the $\mathbf{MS}$ representation is $\ord{N_c^0}$, and the second operator involves the couplings of the axial currents of the core and the excited quark through a second rank tensor $\ell^{(2)}$.
There are four operators at $\ord{1/N_c}$: $M_4$  is a linear combination of two $\ord{N_c^0}$ operators, $M_5$ is a 2-body spin-orbit interaction involving the spin of the core, $M_6$ is the hyperfine core interaction which is known to play the main  role in the spin-flavor symmetry breaking of the masses, $M_7$ is the hyperfine between core and excited quark which in states with $\ell=1$ is expected to give suppressed matrix elements, and finally  $M_8$ is the version of $M_3$ without flavor. The matrix elements of the basis operators were given (in a different normalization) in Table II of  Ref. \cite{Carlson:1998vx}.

For the content of states for $N_c=3$ there are seven possible masses and two mixing angles, which means at most nine independent mass operators.
Up to $\ord{N_c^0}$ there are only three operators, and thus there are six parameter independent relations.
As noted in Ref. \cite{Pirjol:2003ye}, two of these relations are the values of the mixing angles, which are
$\theta_1=\cos^{-1}(-\sqrt{2/3}\,)=2.526$ and  $ \theta_3=\cos^{-1}(-{\sqrt{5/6}}\,)=2.721$ in the present  convention.
It should be noted that using the mass formulas up to, and including $\ord{N_c^0}$ corrections,
one can identify with the nucleon states in the paper of Pirjol and Schat \cite{Pirjol:2003ye} in
the following way: $N_{\frac 12}=N^{T=1}_{\frac 12}$,
$N'_{\frac 12}=N^{T=0}_{\frac 12}$, $N_{\frac 32}=N^{T=2}_{\frac 32}$, $N'_{\frac 32}=N^{T=1}_{\frac 32}$,
where $T$ indicates the tower to which the state belongs as explained in \cite{Pirjol:2003ye}.
Thus, their four alternative results for the $\ord{N_c^0}$ mixing angles (see their Fig. 3) follow
from the four possible assignments Eq. (\ref{sets}) which basically imply shifting each of the angles by $\pi/2$.

The remaining four relations involve only the physical masses and are as follows:
\bea
     M_{N_{\frac{1}{2}}}&=&            M_{N'_{\frac{3}{2}}}   \nonumber\\
     M_{N_{\frac{3}{2}}}&=&            M_{N_{\frac{5}{2}}}  \nonumber\\
     M_{N_{\frac{5}{2}}}-\frac{1}{10}M_{N'_{\frac{1}{2}}}&=&3\; \left(\frac{1}{2} M_{\Delta_{\frac{1}{2}}}-\frac{1}{5} M_{\Delta_{\frac{3}{2}}}\right)\nonumber\\
     M_{\Delta_{\frac{1}{2}}}&=& \frac{1}{6}  M_{N_{\frac{1}{2}}}+ \frac{5}{6}  M_{N_{\frac{5}{2}}}
     \label{eq:LOmassrel}
\eea
Numerically,  these relations are satisfied best with  Set 3 in agreement with the result of Ref. \cite{Pirjol:2003ye} where it was
denoted ``Assignment No.1". However, as observed later at $\ord{1/N_c}$ and after performing the combined fits with decays
and EM amplitudes, a different assignment will be favored. The hyperfine interaction plays the key role  in modifying the LO values of
the mixing angles and in permitting  a different identification of states.

\begin{center}
\begin{figure}[h]
\centerline{\includegraphics[width=14.cm,angle=-0]{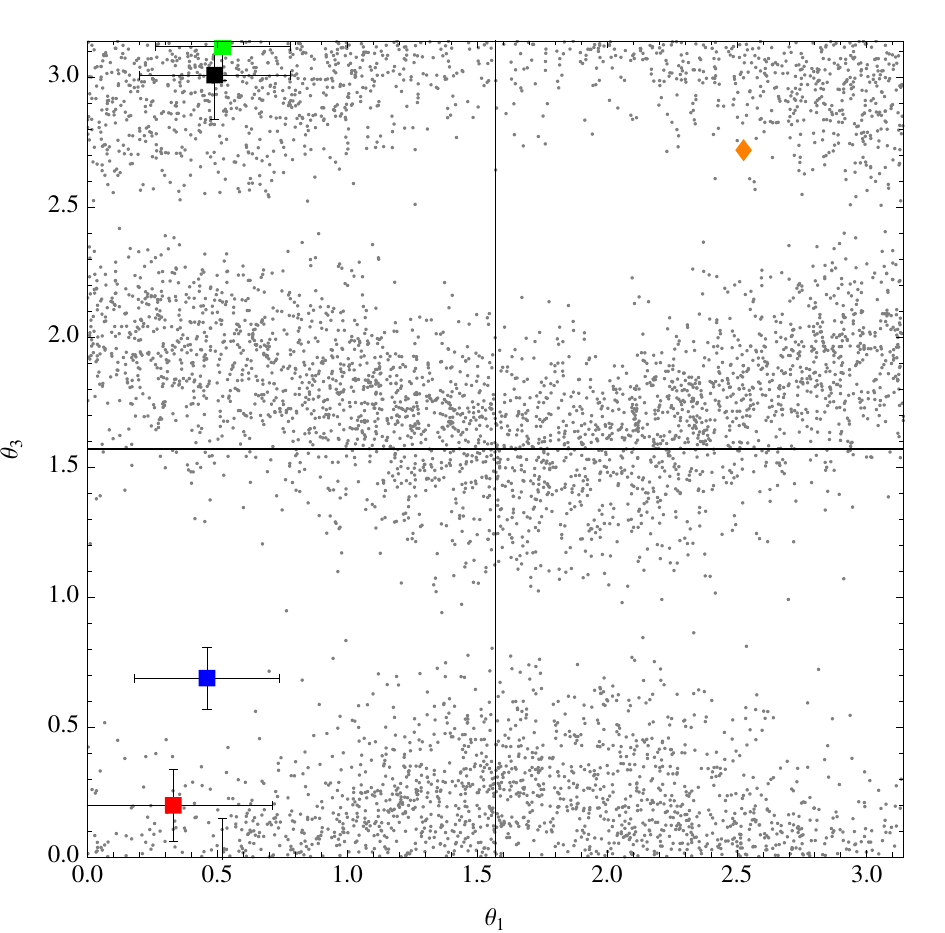}}
\caption{Distribution of possible values of the angles $\theta_1$
and $\theta_3$:  The $\ord{N_c^0}$ angles are depicted by the
orange diamond;  angles consistent with Eq.  (\ref {eq:NLOmassrel})
using the empirical masses and the assignment of states defined in
section [II]  as Set 1 are depicted by the gray distribution of points (for  other assignments shift the
angles by $\pi/2$ accordingly),  and the angles from the global
fits  obtained  in this work  are  depicted in black, red, blue,
and  green  respectively for assignments Set 1 through  4. }
\label{fig:mixingangles}
\end{figure}
\end{center}

Up to $\ord{1/N_c}$ there are eight operators, which means that there will be one mass parameter independent relation which can be checked.
Expressed in terms of the physical masses and mixing angles this relation reads
\bea
\label{eq:NLOmassrel}
&&\!\!\! 3 \left(M_{N_{\frac12}} + M_{N'_{\frac12}} - 4 M_{N_{\frac32}} - 4 M_{N'_{\frac32}} + 6 M_{N_{\frac 52}} + 8  M_{\Delta_{\frac 12}} -
    8  M_{\Delta_{\frac 32}}\right) =  \\[3.mm]
&& \left[13 \cos 2\theta_1 + \sqrt{32} \sin 2\theta_1 \right]  \left(M_{N'_{\frac12}} - M_{N_{\frac12}}\right) -
 4 \left[ \cos 2\theta_3 - \sqrt{20} \sin 2\theta_3 \right] \left(M_{N'_{\frac32}} - M_{N_{\frac32}}\right) .\nonumber
  \eea
  Fig. \ref{fig:mixingangles} shows the correlation between the angles if the assignment of states is the one of Set 1 in Eq. (\ref{sets}).   For the other possible assignments one just shifts the angles by $\frac{\pi}{2}$ correspondingly.

\subsection{Strong decays}
The partial decay widths for the strong decays of the negative parity excited baryons via the emission of a $\pi$ or $\eta$ meson are given by~\cite{Goity:2004ss}:
\begin{equation}
\Gamma^{[\ell_P,I_P]}
= \frac{k_P^{1+2\ell_P}}{8 \pi^2 \Lambda^{2\ell_P}} \frac{M_{B^*}}{M_B} \
\left|\frac{\sum_q C_q^{[\ell_P, I_P]}\
{\cal B}_q(\ell_P,I_P,S,I,J^*,I^*,S^*)}
{\sqrt{(2 J^* + 1)(2 I^*+1)}}\right|^2,
\label{width}
\end{equation}
where the asterisk refers to the  excited baryon, $\ell_P=0$ or 2 is the angular momentum of the meson and $I_P$ its isospin, and $\Lambda$ is an arbitrary  scale introduced for convenience so that all quantities in the sum are dimensionless. By choice in the calculations, $\Lambda=200$ MeV will be taken. ${\cal B}_q(\ell_P,I_P,S,I,J^*,I^*,S^*)$ are reduced matrix elements defined via the Wigner-Eckart theorem as follows:
\bea
& & _{\mbox{\bf S}} \langle  S , S_3 ; I , I_3 \mid
\left( B^{[\ell_P, I_P]}_{[m_P,I_{P_3}]}\right)_q
\mid   J^*, J^*_3 ; I^*, I^*_3  ; S^*  \rangle_{\mbox{\bf MS}}
= \frac{ (-1)^{\ell_P - J^* + S + I_P - I^*+I} }{\sqrt{(2 S + 1)(2 I + 1)}} \times
  \qquad \qquad \nonumber \\
& & \qquad
\langle \ell_P , m_P; J^*, J^*_3 \mid S, S_3  \rangle  \langle I_P, {I_P}_3 ; I^*, I^*_3 \mid I=S , I_3  \rangle
\ {\cal B}_q(\ell_P,I_P,S,I,J^*,I^*,S^*) \ .
\eea
The formulas above hold if the physical states have well defined quark-spin $S^*$. However,  the physical states are a mix of the $S^*=J^*\pm \frac{1}{2}$ quark spin states, and thus they need to be modified in the obvious way via the corresponding mixing of the reduced matrix elements. This of course only happens for the case of the $N_{\frac{1}{2}}$ and $N_{\frac{3}{2}}$ states, and thus the dependence of these partial widths on the angles $\theta_1$ and $\theta_3$.
Note that the analysis is restricted to the S and D-wave decays,  because the  only  possible  G-wave transition   in the decay $N_{\fr52}\to\Delta \pi$ is not yet empirically sorted out, and is also irrelevant to the main discussion of this work. The corresponding bases of baryon decay operators and their associated matrix elements were obtained in Ref. \cite{Goity:2004ss}.
For convenience those operator bases are depicted in Table \ref{strop} of Appendix I. All $\ord{N_c^0}$ operators are 1-body type except for one of the 2-body D-wave for $\pi$ emission. For $S$-wave $\pi$ emission there is one $\ord{N_c^0}$ and three  $\ord{1/N_c}$. For D-wave $\pi$ emission there two $\ord{N_c^0}$ and six  $\ord{1/N_c}$ (two of which are 3-body). For S-wave $\eta$ emission there is one $\ord{N_c^0}$ and one  $\ord{1/N_c}$.

\subsection{Photoproduction helicity amplitudes}

The multipole components of the helicity amplitudes for photo-production of the $[{\bf 20'}, 1^-]$
states  can expressed in terms of the matrix elements of   effective operators as follows \cite{Scoccola:2007sn}:
\begin{eqnarray}
A_\lambda^{ML}\hspace{-0.3cm}& &\hspace{-0.05cm} = \sqrt{\frac{3 \alpha N_c }{4 \omega} }  (-1)^{L+1}\; \eta(B^*) \;
\sum_{n,I} g^{[L,\, I]}_{n,L} (\omega)
\; \left<J^*, \lambda; I^*, I_3;S^*\!\mid
\left({\cal B}_n\right)^{[L,\, I]}_{[1,0]}
\mid \! \frac{1}{2} , \lambda-1 ; \frac{1}{2},  I_{3} \right>\nonumber,
\label{aml2}\\
A_\lambda^{EL}\hspace{-.3cm} & &\hspace{-0.05cm} = \sqrt{\frac{3 \alpha N_c }{4 \omega} } \;(-1)^{L}\; \eta(B^*)
\sum_{ n, I}\; \left[ \sqrt{\frac{L+1}{2L+1}}\ g^{[L,\, I]}_{n,L-1}  (\omega)
                        + \sqrt{\frac{L}{2L+1}}\ g^{[L,\, I]}_{n,L+1}  (\omega)
                        \right]
\nonumber \\ & &  \qquad \qquad\qquad \qquad \times
\left<J^*,\lambda; I^*, I_3;S^*\!\mid
\left({\cal B}_n\right)^{[L,\,
I]}_{[1,0]} \mid \! \frac{1}{2} , \lambda-1 ; \frac{1}{2},  I_{3} \right>,
\label{ael2}
\end{eqnarray}
where $ML$ and $EL$ indicates the respective multipoles. Note that, due to parity conservation,  only
$E1$, $M2$ and $E3$ multipoles are allowed. Moreover, $\lambda=\frac{1}{2},\frac{3}{2}$ is the helicity defined along the $\hat{z}$-axis  which
coincides with the photon momentum, and  $\omega=(M_{B^*}^2-M_N^2)/2 M_{B^*}$ is the
photon energy in the rest frame of the excited baryon $B^*$.
These amplitudes correspond to  the  standard definition as used by the Particle Data Group \cite{Beringer:1900zz},  which includes a sign factor $\eta(B^*)$ that stems from the strong
decay amplitude of the excited baryon to a $\pi N$ state making them independent of the phase
conventions used to define the excited states. The sign factors
\begin{eqnarray}
\eta(B^*) = (-1)^{J^* - \frac{1}{2}}\ \mbox{sign}\left( \langle\ell_\pi N || H_{QCD} || J^* I^* \rangle \right)
\label{sign}
\end{eqnarray}
are on the other hand convention dependent.
The sum over $n$  is over all available basis operators with  the given $[L,I]$ quantum numbers.
The factor $\sqrt{N_c}$ appears as usual for transition matrix elements between excited
and ground state baryons \cite{Goity:2004pw}.
 In the electric multipoles there is  a combination
of the coefficients $g^{[L,\, I]}_{n,L-1}$ and $g^{[L,\,
I]}_{n,L+1}$, and because the operators appearing in these
multipoles do not appear in the magnetic multipoles, one may as
well replace that combination of coefficients by a single term
without any loss of generality. Thus, in what follows  only
  $g^{[L,\, I]}_{n,L-1}$ is kept. These  and the coefficients $g^{[L,\, I]}_{n,L}$
are going to be determined by  fits to the empirical helicity amplitudes. In order to streamline the notation the coefficients will then be denoted by: $XL_n^{(I)}$, $X=E$ or $M$, and $I=0,1$.

As in the case of  decays, one must take into account the effects of state mixing for the physical amplitudes.

The corresponding bases of operators and their associated matrix elements were obtained in Ref. \cite{Scoccola:2007sn}.
For convenience the list is  depicted in Table \ref{helope} of Appendix I.

 As in the case of the masses, which are linear in the effective couplings, there are linear relations for the photo-couplings which are independent of those effective couplings. Those relations at LO which involve  either only neutral or only charged baryons  are displayed in the Appendix II.  Their tests are at this point not significant  because the errors of the empirical amplitudes are still too large.

\section{Global analysis and discussion of  results}

As discussed in the Introduction, the main aim of the present work is
to perform a global and consistent fit of the masses, strong decay
widths an EM helicity amplitudes associated with the low-lying
negative parity baryons.
A summary of the present empirical knowledge of their Breit-Wigner masses, strong decay partial widths
and EM helicity amplitudes for photo-production taken from Ref. \cite{Beringer:1900zz}
is given in Table \ref{tab:empdat}.

In order to perform the fits one proceeds as follows. Starting with a certain set of initial values for the parameters
associated to the mass operators one determines the theoretical masses and mixing angles. Next, with these values
of the mixing angles and a set of values for the strong decay parameters, one obtains  the theoretical predictions for
the strong widths and the strong amplitude signs needed for the calculation of the EM helicity amplitudes, Eq. (\ref{sign}).
Finally, using again the same mixing angles together with the strong signs just determined and a set of values for the helicity
amplitudes parameters, one obtains the theoretical predictions for the EM helicity amplitudes. With all these theoretical
predictions $f_i^{theo}$, and the existing empirical values $f_i^{emp}$ and associated errors summarized in Table \ref{tab:empdat},
the $\chi^2_{dof}=\sum_i[(f_i^{theo}-f_i^{emp})/\Delta f_i^{emp}]^2/n_{dof}$ is calculated. Here, $i=1,\ldots,N_{emp}$ runs over
the different empirically known observables and $n_{dof}=N_{emp}-N_{ip}$, where $N_{ip}$ is the total number of independent
parameters used in the fit. This procedure is iterated using different sets of values for mass, strong decay and EM amplitudes
parameters until a (global) minimum of $\chi^2_{dof}$ is found. The minimization is carried out using the MINUIT program.

\begin{center}
\begin{figure}[!h]
\centerline{\includegraphics[width=14.cm,angle=-0]{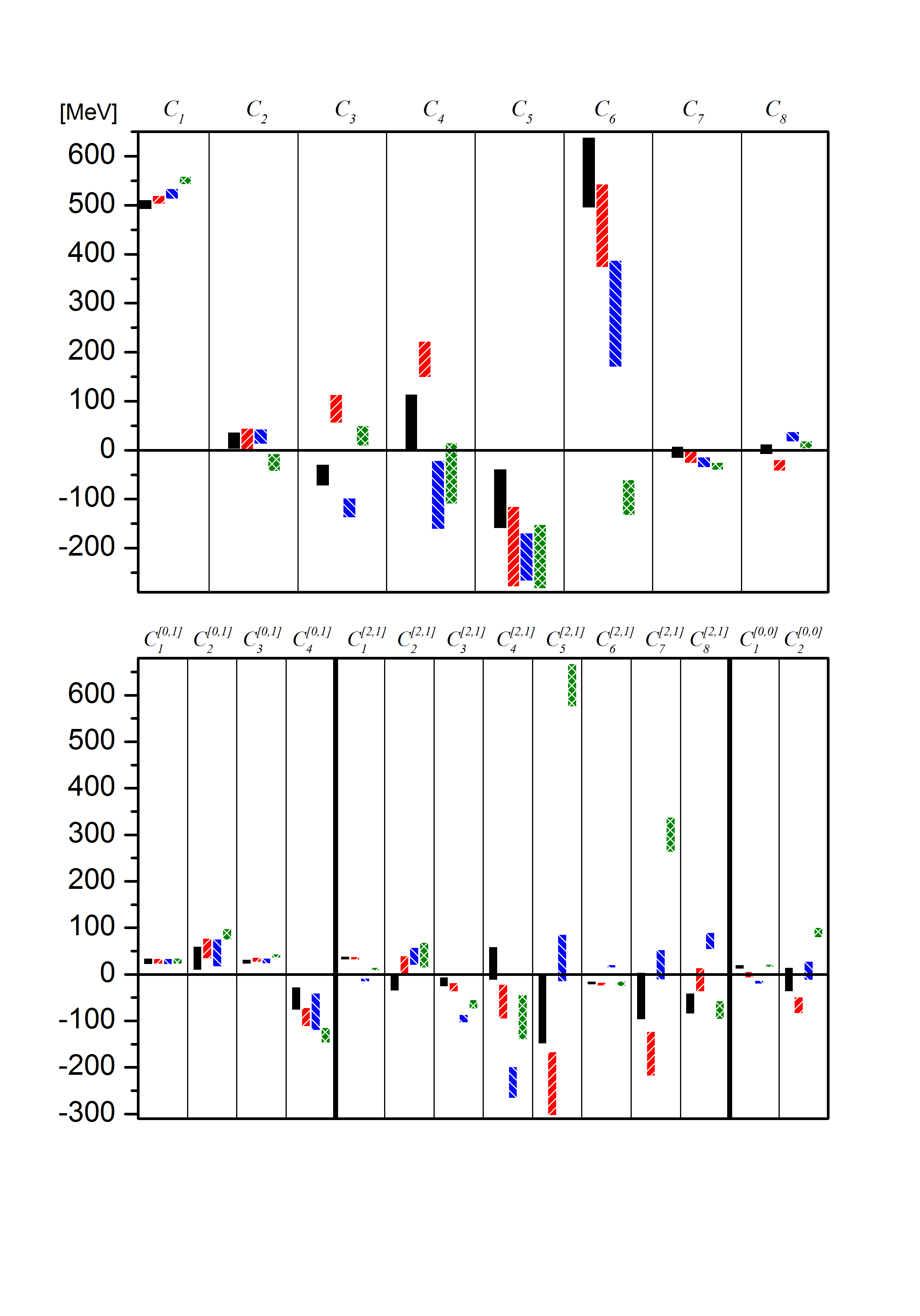}}
\caption{Coefficients with errors from the global fits for the
different assignments of states:   black, red, blue and green
respectively for Sets 1 through 4.}
\label{fig:coefmas}
\end{figure}
\end{center}

\begin{center}
\begin{figure}[!h]
\centerline{\includegraphics[width=14.cm,angle=-0]{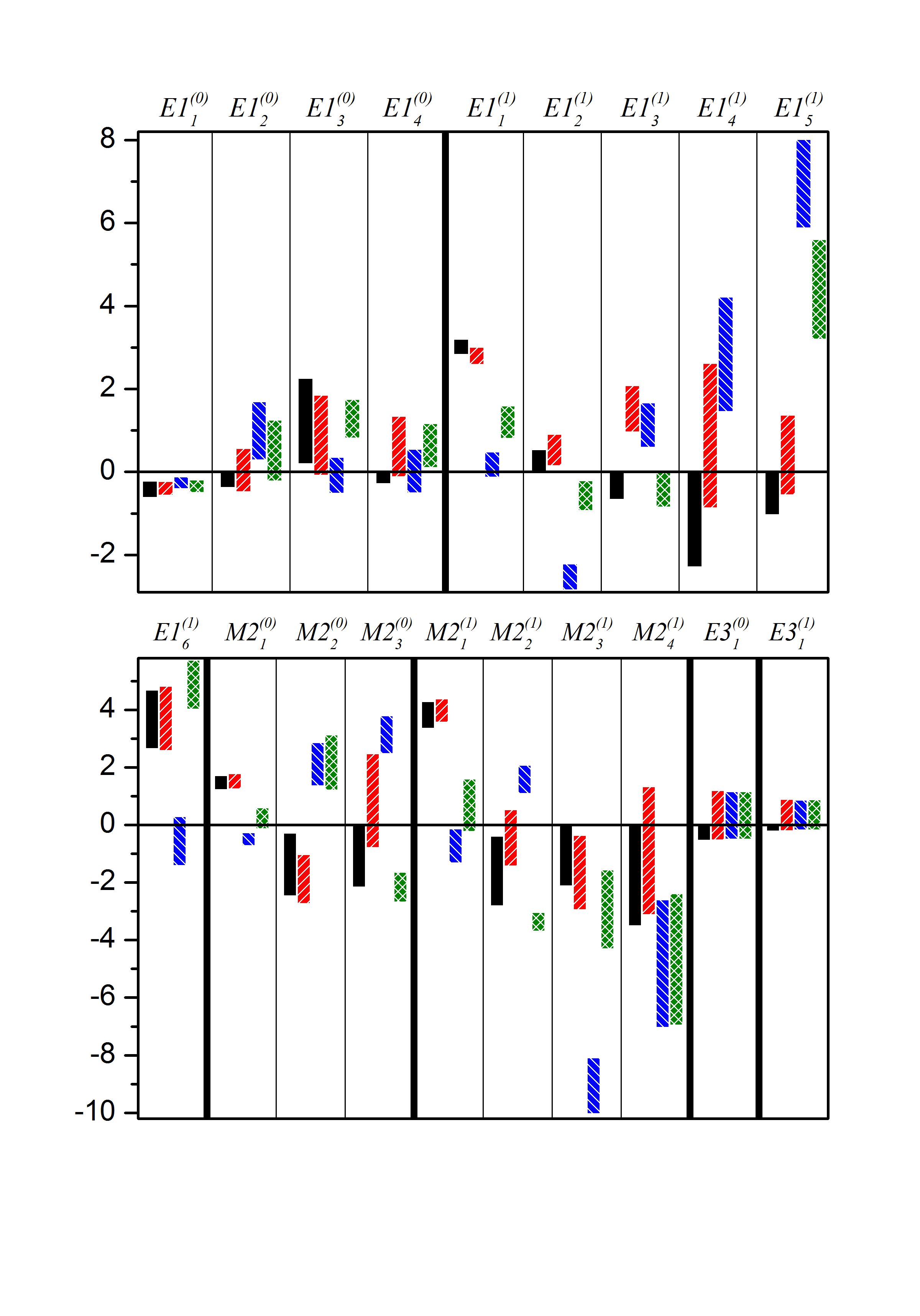}}
\caption{Coefficients with errors from the global fits for the
different assignments of states:   black, red, blue and green
respectively for Sets 1 through 4.}
\label{fig:coefhel}
\end{figure}
\end{center}

Using the procedure mentioned above, fits for the four possible assignments Eq. (\ref{sets}) were performed. It is found
that in all cases it is possible to obtain values of $\chi^2_{dof} < 1$. However, these fits have qualitatively
very different characteristics. This can be observed in Figs. \ref{fig:coefmas} and \ref{fig:coefhel},  which depict the
coefficients associated with mass and strong decay operators and with the EM amplitude operators, respectively.

In order to decide
which assignments are acceptable from a physical point of view one needs to consider a certain number of
criteria. The first one has to do with the naturalness of the coefficients. In fact, if the $1/N_c$ expansion
is assumed to be an appropriate framework,  all the coefficients
should be within a natural range in magnitude. It should be noted that in the case of the strong decays and helicity amplitudes this criterion
should be applied to each partial wave  separately.  A second criterion  is related to the assumption that in the
strong and EM amplitudes, when 1-body operators can contribute they should not be suppressed
with respect to 2-body operators. This assumption is based on the fact that the quark model description
of these amplitudes, where it is assumed by construction  the dominance 1-body operators, is quite successful. Finally, the third criterion
takes into account the importance of the hyperfine interaction in the description of the masses.
In fact, this interaction plays a crucial role in explaining the lightness of the singlet
$\Lambda(1405)$ and $\Lambda(1520)$, whose masses are pushed down by the hyperfine interaction
as shown in \cite{Schat:2001xr}, and as realized in the early paper by Isgur \cite{Isgur:1977ef}. The lightness of those $\Lambda$ baryons is also observed in recent lattice QCD calculation of the baryon spectrum \cite{Edwards:2012fx}, where due to the larger quark masses and the possible suppression of the finite width effects due to the implementation of the lattice calculation, the hyperfine interaction is clearly driving the downward shift of the masses.
The result of the application of these criteria to the results shown in Figs. \ref{fig:coefmas} and \ref{fig:coefhel} is summarized in
Table \ref{criteria}.
From that table one can conclude that the preferred assignment of states is that of Set 1.
This assignment is the one that had been adopted in the early works with the quark model
\cite{Isgur:1977ef,Koniuk:1979vy,Isgur:1978xj} and in the various analysis in the $1/N_c$ expansion,
 \cite{Carlson:1998gw,Carlson:1998vx,Goity:2004ss,Scoccola:2007sn,car94,Pirjol:2003ye}.  The detailed analysis
presented here further establishes that scheme.

The numerical results for the operator coefficients corresponding to the favored  Set 1
are given in Table \ref{coeffit}. Two possibilities have been considered:
Fit 1 includes all the independent operators appearing in the basis of
each observable, while in Fit 2 those operators
whose coefficients are compatible with zero have been excluded.
More specifically, Fit 2 was obtained by taking the result of Fit 1
and removing those operators whose coefficients are compatible with zero
in a subsequent way. In the case that at certain step there were more than
one operator in this situation, those with higher relative error were removed.

It is interesting to compare the operator coefficients obtained in the present
global analysis with those reported in
Refs. \cite{Carlson:1998vx,Goity:2004ss,Scoccola:2007sn}
where fits were made independently for each baryon property.
In making this comparison one should keep in mind that some new
empirical values became available since those analysis were performed.
In the case of strong decays it is  observed that except for a few exceptions
the coefficients obtained in global and non-global analysis are quite
similar when error bars are taken  into account. One of the exceptions
concerns the coefficients $C^{[0,1]}_i$ with $i=2,3,4$ which showed a
degeneracy in the non-global fit. Such degeneracy is removed in the
present global analysis. The other is related to the coefficients
$C^{[2,1]}_{7,8}$ associated to 3-body operators which were not taken
into account in Ref. \cite{Goity:2004ss}. While the present analysis
indicates that $C^{[2,1]}_{7}$ is indeed compatible with zero,
$C^{[2,1]}_{8}$ turns out to be comparable
to the coefficient of the corresponding 1-body operator. On the other hand,
note that $C^{[2,1]}_{2}$ which was non-vanishing in the previous analysis
turns to be compatible with zero here. Finally,
$C^{[0,0]}_2$ which was not included in non-global analysis
is also compatible with zero here.
Making a comparison between the present and previous works on
EM decays \cite{Scoccola:2007sn}, one notices that the values
of the coefficient do not change appreciably. In most cases the average
values are very similar, and they overlap within errors.
It is interesting to revisit the Moorhouse rule \cite{Moorhouse}, which states that the EM transitions of proton excited states with quark spin $S^*=\frac{3}{2}$ are suppressed. An explanation of the rule in the quark model stems from coupling the photon to the constituent quarks with the isoscalar/isovector strength determined by the quark charges. State mixing and the presence of 2-body operators give violations to the rule. The Table \ref{moor} gives the quark-spin content of the proton states as determined from the mixing  angles obtained in the fits to Set 1 and a check if the rule is closely satisfied. This check represents a good test of the identification of states corresponding to the Set 1. For Sets 3 and 4   the only way to make the suppressions work is through extra strength in the 2-body operators as shown in Fig. \ref{fig:coefhel}, which is expected to be  unnatural. The case of $p(1675)$ is a particularly unambiguous test as it is unaffected by mixings  \cite{Moorhouse}.

In the strong decays the following selection rules result at LO, i.e., when the mixing angles are the ones obtained in the expansion of the masses up to $\ord{N_c^0}$ and at LO in the decay amplitudes (1-body operators):   for  the  S-wave decays $N_{\frac{1}{2}}\nrightarrow \eta N$, $N'_{\frac{1}{2}}\nrightarrow \pi N$,  $N_{\frac{3}{2}}\nrightarrow \pi\Delta$,  and for  the  D-wave decay $N'_{\frac{1}{2}}\nrightarrow \pi\Delta$.   These can be checked using the Tables III, IV and V of Ref. \cite{Goity:2004ss}.
Since the mixing angle $\theta_1$ changes by more than $\pi/2$ from LO to NLO, one expects that for the $J=1/2$ nucleons the selection rules will be poorly satisfied. In fact the decays $N_{\frac{1}{2}}\to \eta N$, $N'_{\frac{1}{2}}\to  \pi N$ are not suppressed at all, while the D-wave decay $N'_{\frac{1}{2}}\to \pi\Delta$ is quite suppressed with respect to the S-wave one, $N'_{\frac{1}{2}}\to \pi N$, but this is simply because of being a higher partial wave. The large change in the angle $\theta_1$ from LO to NLO is due to the fact that the mass operators of $\ord{N_c^0}$ are quite weak and the NLO hyperfine mass operator is very strong, leading to a large rearrangement in the composition of the $J=1/2$ nucleon states.  Finally the suppression of the S-wave  decay  $N_{\frac{3}{2}}\to \pi\Delta$ should be well fulfilled because the angle $\theta_3$ does not change too much from LO to NLO. This seems to be indeed the case as the D-wave decay $N_{\frac{3}{2}}\to \pi\Delta$ has similar rate to that S-wave decay.

\section{CONCLUSIONS}

The determination of the spin-flavor structure of the negative parity baryons is of fundamental importance in the study of excited baryons, as it serves to unravel the QCD dynamics responsible for their properties. Of particular importance is the structure of the $J=1/2$ and 3/2 nucleons, which coming in pairs can mix. The mixing is a sensitive measure of that dynamics.  The connection between the theoretical states and the observed ones is a priori a four-fold undetermined problem. The physics is definitely different in each case, as it has been shown in this work through the global analysis of observables. The study comes to the conclusion that there is a favored assignment, based on the three criteria outlined in section III.  Set 1 is acceptable on all counts, while Set 2 comes second, failing only in one criterion, namely  that 2-body operators dominate over 1-body ones  in the strong decays.  Set 1 is moreover the favored one in the simplest versions of the quark model with suppressed spin-orbit interaction, which indicates that the seemingly simplest dynamics is the most realistic one.
Concerning the mixing angles, this work finds $(\theta_1, \theta_3) = (0.49 \pm 0.29  , 3.01 \pm 0.17)$ and $(0.40 \pm 0.13  , 2.96 \pm 0.05)$ for Fits 1 and 2 respectively. For comparison
in a previous non-global analysis of the strong decays using the $1/N_c$ \cite{Goity:2004ss},
the value  $\theta_1 = 0.39 \pm 0.11$ was obtained and $\theta_3$ exhibited a two-fold degeneracy $\theta_3 = (2.38,2.82) \pm 0.11$. In this sense the present global analysis
is important to remove such degeneracy in the values for $\theta_3$.
It is useful to compare the results obtained in the present analysis with old results obtained with analyses based on the quark model and of the $SU(6)_W$  approach. The present values
are in good agreement with the earlier quark model determination $(\theta_1, \theta_3)=(0.55,\;3.03)$, obtained  from the analysis of strong decays alone \cite{Isgur:1978xj}.  The present results for $\theta_3$ agree with old results in the $SU(6)_W$ approach \cite{lll}, while for $\theta_1$ those references disagreed with each other, for the most part because the input partial widths for the $N_{\frac 12}$ baryons had changed in the meantime,  with the latest agreeing with the present result.
Finally, it is worthwhile to stress that further improvement in the empirical inputs for partial decay widths and helicity amplitudes is necessary in order to reduce the mixing angles uncertainties and to establish with an even stronger emphasis  Set 1 as the correct assignment of spin-flavor states.

\section*{ACKNOWLEDGEMENTS}
This work was supported by DOE Contract No. DE-AC05-06OR23177 under which JSA operates the Thomas Jefferson National Accelerator Facility, and by the National Science Foundation (USA) through grants PHY-0855789 and PHY-1307413 (J.~L.~G.).
This work has been partially funded by CONICET (Argentina) under grants
PIP 00682 and by ANPCyT (Argentina) under grant PICT-2011-0113 (E.~G.~U. and N.~N.~S.).

\section*{APPENDIX I: BASES OF OPERATORS}

For the reader's convenience   the  operator  bases are shown: Table \ref{masop} gives the mass operators, Table \ref{strop} operators
for the strong decays and Table \ref{helope} operators for the helicity amplitudes. They were respectively obtained in Refs.
\cite{Carlson:1998vx,Goity:2004ss,Scoccola:2007sn}, where the corresponding matrix elements relevant
for the present work can be also found.

\section*{APPENDIX II: PARAMETER INDEPENDENT RELATIONS FOR  HELICITY AMPLITUDES}

The parameter independent relations satisfied at LO  by the photo-couplings are shown below. At LO the relations are given separately for the neutral and unit charge baryons. Other LO relations involving simultaneously neutral and charged baryons are not given here, but can be  easily obtained. In the following the notation $s_i\equiv \sin\theta_i$,  $c_i\equiv \cos\theta_i$ is used.

Charged baryon relations:
\beq
\begin{array}{l}
\begin{array}{l}
\left(\sqrt{\frac{2}{3}}  \,c_1  \,c_3 +
     2 \sqrt{\frac{6}{5}} \,c_1 \,s_3 +
      \frac{2}{\sqrt{15}} \,s_1 \,s_3\right)A^{p}_{\frac{1}{2}}[N_{\frac{1}{2}}]
+
\frac{1}{\sqrt{3}} A^{p}_{\frac{1}{2}}[N_{\frac{3}{2}}]
+
A^{p}_{\frac{3}{2}}[N_{\frac{3}{2}}]
\\ [4.mm]
+ \left(\frac{2}{\sqrt{15}}  \,c_1  \,s_3 -
      \sqrt{\frac{6}{5}} \,s_1 \,s_3 -
      \sqrt{\frac{2}{3}} \,s_1 \,c_3\right)A^{p}_{\frac{1}{2}}[N'_{\frac{1}{2}}]
+\left(\frac{2}{9} \left(\sqrt{10}+2 \sqrt{15}\right) \,s_3-\frac{1}{\sqrt{3}}\,c_3\right)
A^{N}_{\frac{1}{2}}[\Delta_{\frac{1}{2}}]
\\ [4.mm]
- \left(\frac{1}{\sqrt{6}}\,c_3+\frac{10+19 \sqrt{6}}{18 \sqrt{5}}\,s_3\right)
A^{N}_{\frac{1}{2}}[\Delta_{\frac{3}{2}}] -
\left(\frac{1}{\sqrt{2}}\,c_3+\frac{57+5 \sqrt{6}}{9 \sqrt{10}}\,s_3\right)
A^{N}_{\frac{3}{2}}[\Delta_{\frac{3}{2}}]
=0
\end{array}
%
\\\\
%
\begin{array}{l}
 \left(2 \sqrt{\frac{6}{5}}\,c_1\,c_3 + \frac{2}{\sqrt{15}} \,s_1\,c_3 - \sqrt{\frac{2}{3}} \,c_1\,s_3\right)
 A^{p}_{\frac{1}{2}}[N_{\frac{1}{2}}]
 +\left(\frac{2}{\sqrt{15}}\,c_1\,c_3 -2 \sqrt{\frac{6}{5}} \,s_1\,c_3 + \sqrt{\frac{2}{3}} \,c_1\,s_3\right)
A^{p}_{\frac{1}{2}}[N'_{\frac{1}{2}}]
\\ [4.mm]
+ \frac{1}{\sqrt{3}} A^{p}_{\frac{1}{2}}[N'_{\frac{3}{2}}]
 +A^{p}_{\frac{3}{2}}[N_{\frac{3}{2}}]
 +\left(\frac{2}{9} \left(\sqrt{10}+2  \sqrt{15}\right) {\,c_3}+\frac{1}{\sqrt{3}} \,s_3\right)
 A^{N}_{\frac{1}{2}}[\Delta_{\frac{1}{2}}]
\\ [4.mm]
 + \left(\frac{1}{\sqrt{6}}\,s_3-\frac{10+19 \sqrt{6}}{18 \sqrt{5}}\,c_3\right)
 A^{N}_{\frac{1}{2}}[\Delta_{\frac{3}{2}}]+
 \left(\frac{1}{\sqrt{2}}\,s_3-\frac{57 \sqrt{10}+10\sqrt{15}}{90}   \,c_3\right)
A^{N}_{\frac{3}{2}}[\Delta_{\frac{3}{2}}]
 =0
\end{array}
%
\\\\
%
\begin{array}{l}
 \frac{1}{3\sqrt{5}}\left(8 \,c_1+\frac{1}{\sqrt{2}} \,s_1\right)
A^{p}_{\frac{1}{2}}[N_{\frac{1}{2}}]
+\frac{\sqrt{2}}{3} \left(2 \sqrt{5}  \,c_3+ s_3 \right)
A^{p}_{\frac{1}{2}}[\Delta_{\frac{1}{2}}]
+
\frac{1}{\sqrt{2}} A^{p}_{\frac{1}{2}}[N_{\frac{5}{2}}]
+
A^{p}_{\frac{3}{2}}[N_{\frac{5}{2}}]
\\ [4.mm]
+\frac{1}{3\sqrt{5}}\left(\frac{1}{\sqrt{2}} \,c_1 - 8 \,s_1\right)
A^{p'}_{\frac{1}{2}}[N_{\frac{1}{2}}]
+\frac{\sqrt{2}}{3}  \left( \,c_3-2 \sqrt{5} \,s_3 \right)
A^{p'}_{\frac{1}{2}}[\Delta_{\frac{1}{2}}]
\\ [4.mm]
+ \frac{2  \sqrt{15}-3 \sqrt{10} }{54}
A^{N}_{\frac{1}{2}}[\Delta_{\frac{1}{2}}]
+  \frac{663-5 \sqrt{6}} {108 \sqrt{5}}
A^{N}_{\frac{1}{2}}[\Delta_{\frac{3}{2}}]
-\frac{ 5 \sqrt{10}+139 \sqrt{15} }{180}
A^{N}_{\frac{3}{2}}[\Delta_{\frac{3}{2}}]
 =0
\end{array}
\end{array}
\eeq
Neutral baryon relations:
\beq
\begin{array}{l}
\begin{array}{l}
 \left(\sqrt{\frac{2}{3}} \,c_1 \,c_3+2\sqrt{\frac{6}{5}} \,c_1\,s_3+ \frac{2}{\sqrt{15}}\,s_1\,s_3\right)
 A^{n}_{1/2}[N_{1/2}]
+
\frac{1}{\sqrt{3}}
A^{n}_{1/2}[N_{3/2}]
+
A^{n}_{1/2}[N_{3/2}]
\\ [4.mm]
+\left(\frac{2}{\sqrt{15}}\,c_1 \,s_3 - \sqrt{\frac{2}{3}} \,s_1\,c_3 - 2\sqrt{\frac{6}{5}}\,s_1\,s_3\right)
A^{n}_{1/2}[N'_{1/2}]
+\left(\frac{1}{\sqrt{3}}\,c_3-\frac{2\sqrt{5}}{9}\left(\sqrt{2}+2\sqrt{3}\right)\,s_3\right)
A^{N}_{1/2}[\Delta_{1/2}]
\\ [4.mm]
+\left(\frac{1}{\sqrt{6}}\,c_3+\frac{\sqrt{5}}{90}\left(10+19\sqrt{6}\right)\,s_3\right)
A^{N}_{1/2}[\Delta_{3/2}]
+\left(\frac{1}{\sqrt{2}}\,c_3+\frac{\sqrt{5}}{90}\left(57\sqrt{2}+10\sqrt{3}\right)\,s_3\right)
A^{N}_{3/2}[\Delta_{3/2}]=0
\end{array}
\\\\
\begin{array}{l}
\left(2\sqrt{\frac{6}{5}} \,c_1 \,c_3 + \frac{2}{\sqrt{15}} \,s_1\,c_3 - \sqrt{\frac{2}{3}} \,c_1\,s_3\right)
A^{n}_{1/2}[N_{1/2}]
+ \left(\frac{2}{\sqrt{15}} \,c_1 \,c_3 - 2\sqrt{\frac{6}{5}} \,s_1\,c_3 + \sqrt{\frac{2}{3}} \,s_1 \,s_3\right)
A^{n}_{1/2}[N'_{1/2}]
\\ [4.mm]
+ \frac{1}{\sqrt{3}}
A^{n}_{1/2}[N'_{3/2}]
+
A^{n}_{3/2}[N'_{3/2}]
-\left( \frac{2}{9}\left(\sqrt{10}+2 \sqrt{15}\right)\,c_3 +\frac{1}{\sqrt{3}}\,s_3\right)
A^{N}_{1/2}[\Delta_{1/2}]
\\ [4.mm]
+\left(\frac{10\sqrt{5}+19 \sqrt{30}}{90}\,c_3-\frac{1}{\sqrt{6}}\,s_3\right)
A^{N}_{1/2}[\Delta_{3/2}]+\left(\frac{57\sqrt{10}+10\sqrt{15}}{90}\,c_3-\frac{1}{\sqrt{2}}\,s_3\right)
A^{N}_{3/2}[\Delta_{3/2}]=0
\end{array}
\\\\
\begin{array}{l}
\frac{1}{3\sqrt{10}}\left( 8\sqrt{2}\,c_1+ s_1\right)
A^{n}_{1/2}[N_{1/2}]
+\frac{\sqrt{2}}{3} \left(2 \sqrt{5} \,c_3+ s_3 \right)
A^{n}_{1/2}[N_{3/2}]
+
\frac{1}{\sqrt{2}}
A^{n}_{1/2}[N_{5/2}]
\\ [4.mm]
+
A^{n}_{3/2}[N_{5/2}]
+\frac{1}{3\sqrt{10}}\left( c_1 - 8\sqrt{2} \,s_1\right)
A^{n}_{1/2}[N'_{1/2}]
+\frac{\sqrt{2}}{3} \left( c_3 - 2 \sqrt{5} \,s_3 \right)
A^{n}_{1/2}[N'_{3/2}]
\\ [4.mm]
+\frac{10 \left( 3 \sqrt{2}-2 \sqrt{3}\right)}{108 \sqrt{5}}
A^{N}_{1/2}[\Delta_{1/2}]
+ \frac{-663+5 \sqrt{6}}{108 \sqrt{5}}
A^{N}_{1/2}[\Delta_{3/2}]
+ \frac{ 3 \left(5 \sqrt{2}+139 \sqrt{3}\right)}{108 \sqrt{5}}
A^{N}_{3/2}[\Delta_{3/2}]
   =0
\end{array}
\end{array}
\eeq

\newpage

\begin{center}
\begin{table}[h!]
\hspace*{-.5cm}
\renewcommand{\arraystretch}{1.2}
\begin{tabular}{cccccccccc} \hline \hline
\hspace*{3mm}State\hspace*{3mm} & \hspace*{3mm} Mass \hspace*{3mm} &\hspace*{3mm} Width\hspace*{3mm}
& \multicolumn{2}{c}{Branching ratio [\%]}
&\multicolumn{2}{c}{ \hspace*{3mm} $A^{N}_{\frac{1}{2}}$}
&\multicolumn{2}{c}{$A^{N}_{\frac{3}{2}}$} &   \\
&&& S-wave & D-wave &
    $A^{p}_{\frac{1}{2}}$  &$A^{n}_{\frac{1}{2}}$  & $A^{p}_{\frac{3}{2}}$  &$A^{n}_{\frac{3}{2}}$
    &              \\[3.mm]
&[MeV]&[MeV]&        &
&& \multicolumn{2}{c}{ [$10^{-3} \times$ GeV$^{-\frac{1}{2}}$]        }
 \\[3.mm]\hline
$N(1535)$ &  $1538(18)$    & $150(25) $ &
      $\pi N: 45(10)$              &     $\pi \Delta:2(2)$  &90(25)&-46(27)&&&      \\
&&&  $\eta N: 42(10)$         &                                 &&&&   \\
$N(1520)$&  $1523(8)$   & $113(13)$ &
  $\pi \Delta: 15(5)$           &        $\pi N:  60(5)$        &-24(9)&-59(9)&166(5)&-139(11)& \\
&&&                                 &     $\pi \Delta: 13(3)$        &&&&&  \\
&&&                                 &  $\eta N:~ {\rm unknown}$      &&&&    \\
$N(1650)$ &  $1660(20)$ & $150(30)$  &
      $\pi N: 70(20)$          &        $\pi \Delta: 13(13)$      &53(16)&-15(21)&&& \\
&&&   $\eta N: 10(5)$          &                                  &&&&&  \\
$N(1700)$ &  $1700(50)$ & $175(75)$ &
  $\pi \Delta: ~ 50(40)$       &   $\pi N:~      12(5)     $    &-18(13)&0(50)&-2(24)&-3(44)& \\
&&&                                 &  $\pi \Delta: ~ 10(10)$    &&&&  \\
&&&                                 &     $\eta N:~ {\rm unknown}$    &&&&   \\
$N(1675)$ &   $1678(8) $ & $148(18)$  &
                                     &   $\pi N:~      40(5)$          &19(8)&-43(12)&15(9)&-58(13)& \\
&&&                                 &   $\pi \Delta: ~ 50(15)$    &&&&& \\
&&&                                 &       $\eta N:~ {\rm unknown}$    &&&&& \\
$\Delta$(1620)& $1645(30)$&  $143(8)$ &
  $\pi N:~ 25(5)$                  &   $\pi \Delta: 45(15)$         &\multicolumn{2}{c}{27(11)}&&&  \\
&&&           ~                     &       $\eta \Delta$:    {\rm no~ phase}              &&&&& \\[-5mm]
&&&              &     space        &&&&  \\
$\Delta$(1700)& $1720(50)$& $300(100)$ &
   $\pi \Delta: 38(13)$        &       $\pi N:~      15(5)$      &\multicolumn{2}{c}{104(15)}&\multicolumn{2}{c}{85(22)}& \\
&&&   $\eta \Delta$: no phase             &   $\pi \Delta: 10(5)$          &&&&  \\[-5mm]
&&&   space            &            &&&&  \\[-5mm]
&&&                                 &     $\eta \Delta$:   {\rm no~ phase}                  &&&& \\[-5mm]
&&&              &     space        &&&&  \\ [2mm] \hline \hline
\end{tabular}\caption{\footnotesize Non-strange negative parity baryons with their partial decay widths and helicity photoproduction amplitudes
as given by the PDG \cite{Beringer:1900zz}. }
\label{tab:empdat}
\end{table}
\end{center}

\begin{center}
\begin{table}[htdp]
\begin{tabular}{cccccc}\hline\hline
Set & ~~~~~~~~~~~1 & 2 & 3 & 4\\ \hline
Naturalness &
$\begin{array}{lll}
M & &\checkmark \\
D & &\checkmark\\
EM &&\checkmark \end{array}$ &
$\begin{array}{ll}
 &\checkmark \\
 &$ \checkmark $ \\
 &\checkmark \end{array} $&
$\begin{array}{ll}
 &$\xmark $ \\
 & $\xmark $\\
 &\checkmark \end{array} $&
$\begin{array}{ll}
 &\checkmark \\
 &$\xmark $\\
 &$\xmark $ \end{array} $\\ \hline
1-B dominance &
$\begin{array}{lll}
D & &\checkmark\\
EM &&\checkmark \end{array}$ &
$\begin{array}{ll}
  &$ \xmark $\\
 &\checkmark \end{array} $&
$\begin{array}{ll}
 &$\xmark $\\
 &$\xmark $ \end{array}$ &
$\begin{array}{ll}
 &  $\xmark $\\
 & $\xmark $ \end{array} $\\ \hline
HF term ~~~& M~~~~~ \checkmark & \checkmark & \checkmark  & ~  \xmark
\\ \hline\hline
\end{tabular}
\caption{Summary of  application of the criteria discussed in the text to results shown in Figs. \ref{fig:coefmas} and \ref{fig:coefhel}. $M$, $D$ and $EM$ refer to masses, decays and EM helicity amplitudes respectively.}
\label{criteria}
\end{table}
\end{center}

\begin{center}
\begin{table}[!h]
\renewcommand{\arraystretch}{1.06}
\begin{tabular}{ccccccccccccc}
\hline\hline
\multicolumn{3}{c}{Mass [MeV]}                     &$\quad$ &  \multicolumn{3}{c}{Strong decays}            &$\quad$   &  \multicolumn{3}{c}{EM helicity amplitudes }\\
                     &\hspace*{2mm} Fit 1 \hspace*{2mm}  & \hspace*{2mm} Fit 2  \hspace*{2mm}   & &
                     &\hspace*{2mm} Fit 1 \hspace*{2mm}  & \hspace*{2mm} Fit 2  \hspace*{2mm}   & &
                     &\hspace*{2mm} Fit 1 \hspace*{2mm}  & \hspace*{2mm} Fit 2  \hspace*{2mm}    \\
\cline{1-3}\cline{5-7}\cline{9-11}
    $C_1 $           &497(5)          &497(3)      & &$C^{[0,1]}_1 $   &  23(2)  &22(2)          &  &$E1^{(0)}_1$&-0.4(0.2)&  -0.4(0.2)\\
    $C_2 $           &20(17)          &28(11)      & &$C^{[0,1]}_2 $   &  35(27)     &25(12)     &  &$E1^{(0)}_2$&0.2(0.6) &    -        \\
    $C_3 $           &-50(22)         &-48(10)     & &$C^{[0,1]}_3 $   &  27(6)  &26(5)          &  &$E1^{(0)}_3$&1(1)   & 0.7(0.7)  \\
    $C_4 $           &57(58)          &32(27)      & &$C^{[0,1]}_4 $   &  -52(26)    &-41(16)    &  &$E1^{(0)}_4$&0.4(0.7) &     -       \\
    $C_5 $           &-99(61)         &-98(32)     & &$C^{[2,1]}_1 $   &  3.6(0.2) &3.6(0.2)     &  &$E1^{(1)}_1$&3.0(0.2) & 3.1(0.2)  \\
    $C_6$            &567(72)         &577(41)     & &$C^{[2,1]}_2 $   &  -2(2)  &     -         &  &$E1^{(1)}_2$&0.3(0.3) & 0.3(0.2)  \\
    $C_7$            &-4(13)          &     -      & &$C^{[2,1]}_3 $   &  -2(1)  &-2(1)          &  &$E1^{(1)}_3$&-0.1(0.6)&     -       \\
    $C_8$            &3(11)           &     -      & &$C^{[2,1]}_4 $   &  2(4)   &     -         &  &$E1^{(1)}_4$&0(2)  &     -       \\
                     &                &            & &$C^{[2,1]}_5 $   &  -8(8)  &     -         &  &$E1^{(1)}_5$&-0.1(0.9)&     -       \\
\cline{1-3}
\multicolumn{3}{c}{Mixing angles}            & &$C^{[2.1]}_6 $   &  -1.6(0.2)&  -1.5(0.2)  &  &$E1^{(1)}_6$&4(1)   & 3.8(0.7)  \\
                     &   Fit 1        &  Fit 2     & &$C^{[2,1]}_7 $   &  -5(5)  &  -            &  &$M2^{(0)}_1$&1.5(0.3) &  1.5(0.3) \\
\cline{1-3}
  $\theta_1 $        &   0.49(0.29)   &0.40(0.13)  & &$C^{[2,1]}_8 $   &  -6(2)  &  -7(2)        &  &$M2^{(0)}_2$&-1(1)  &  -1.6(0.8)\\
  $\theta_3 $        &   3.01(0.17)   &2.96(0.05)  & &$C^{[0,0]}_1 $   &  16(6)  &  18(2)        &  &$M2^{(0)}_3$&0(2)  &     -       \\
                     &                &            & &$C^{[0,0]}_2 $   & -11(27)     &    -      &  &$M2^{(1)}_1$&3.8(0.5) & 3.8(0.4)  \\
\cline{1-3}
                     &   Fit 1        &  Fit 2     & &                 &             &           &  &$M2^{(1)}_2$&-1.6(1.2)  & -2.2(0.6) \\
\cline{1-3}
$\chi^2_{\rm dof}$   &  0.39          &   0.34     & &                 &             &           &  &$M2^{(1)}_3$&-1(1)  &     -       \\
${\rm dof}$          &   5            &    22      & &                 &             &           &  &$M2^{(1)}_4$&-1(2)  &     -       \\
                     &                &            & &                 &             &           &  &$E3^{(0)}_1$&0.3(0.9) &     -       \\
                     &                &            & &                 &             &           &  &$E3^{(1)}_1$&0.3(0.6) &     -      \\[2mm]\hline
\hline
\end{tabular}
\caption{Results for the fits to Set 1.  The parameters fitted are those  associated with the basis operators for masses, strong decays and helicity amplitudes. The parameters for the latter two are dimensionless.
The results for two fits are given. In Fit 1 all operators are included while in Fit 2  operators with coefficients compatible with zero are
disregarded.} \vspace*{2mm}
\end{table}
\label{coeffit}
\end{center}

\begin{center}
\begin{table}[!h]
\renewcommand{\arraystretch}{.903}
\begin{tabular}{cccccccc}
\hline\hline
 \multicolumn{8}{c}{Masses [MeV]} \\
  \multicolumn{2}{c}{State}                &&\hspace*{.5cm}  Empirical   \hspace*{.5cm}  && \hspace*{.5cm}Fit 1  \hspace*{.5cm}     && \hspace*{.5cm} Fit 2 \hspace*{.5cm} \\
\hline
 \multicolumn{2}{c}{$N(1535)$}             && 1538(18) &&1539  && 1539\\
 \multicolumn{2}{c}{$N(1520)$}             && 1523(8)  &&1523  && 1523\\
 \multicolumn{2}{c}{$N(1650)$}             && 1660(20) &&1659  && 1658\\
 \multicolumn{2}{c}{$N(1700)$}             && 1700(50) &&1697  && 1718\\
 \multicolumn{2}{c}{$N(1675)$}             && 1678(8)  &&1678  && 1678\\
 \multicolumn{2}{c}{$\Delta(1620)$}        && 1645(30) &&1651  && 1646\\
 \multicolumn{2}{c}{$\Delta(1700)$}        && 1720(50) &&1703  && 1701\\ [3.mm]
\hline\hline
\multicolumn{8}{c}{Strong partial decay widths [MeV]}                                               \\
 \multicolumn{2}{c}{Channel}               && Empirical      &&  Fit 1      &&  Fit 2                \\
\hline
    & $N(1535)\rightarrow \pi N$           &&68(19)    &&61.5   &&61.4            \\
    & $N(1520)\rightarrow \pi \Delta$      &&17(6)     &&15.0   &&14.6            \\
$\pi$
    & $N(1650)\rightarrow \pi N$           &&105(37)   &&95.4   &&88.3            \\
S-wave& $N(1700)\rightarrow \pi \Delta$    &&88(79)    &&133  &&149          \\
    & $\Delta(1620)\rightarrow \pi N$      &&35(7)     &&35.6   &&36.3            \\
    &$\Delta(1700)\rightarrow\pi\Delta$    &&113(53)   &&137  &&128           \\
\hline
    & $N(1535)\rightarrow \pi \Delta$      &&3(3)      &&2.0    &&2.02             \\
    & $N(1520)\rightarrow \pi    N $       &&68(9)     &&66.0   &&72.6            \\
    & $N(1520)\rightarrow \pi \Delta$      &&14(3)     &&12.4   &&9.36             \\
    & $N(1650)\rightarrow \pi \Delta$      &&19(19)    &&22.8   &&19.5            \\
$\pi$
    & $N(1700)\rightarrow\pi N$            &&21(13)    &&20.5   &&26.0            \\
D-wave & $N(1700)\rightarrow\pi\Delta$     &&18(19)    &&20.4   &&24.2            \\
    & $N(1675)\rightarrow\pi N$            &&59(10)    &&59.7   &&57.7            \\
    & $N(1675)\rightarrow\pi\Delta$        &&73(24)    &&82.0   &&58.2            \\
    & $\Delta(1620)\rightarrow\pi\Delta$   &&64(22)    &&75.3   &&74.1             \\
    & $\Delta(1700)\rightarrow\pi N$       &&45(15)    &&44.7   &&47.0             \\
    & $\Delta(1700)\rightarrow\pi\Delta$   &&30(18)    &&33.5   &&30.2             \\
\hline
$\eta$
    & $N(1535)\rightarrow \eta N$          &&63(18)    &&63.0   &&65.3             \\
S-wave
    & $N(1650)\rightarrow \eta N$          &&15(8)     &&15.0   &&12.6             \\
\hline \hline
\end{tabular}
\caption{ Results for masses and partial decay widths from the fits to Set 1. }
\label{resfita}
\end{table}
\end{center}

\begin{center}
\begin{table}[!h]
\renewcommand{\arraystretch}{.903}
\begin{tabular}{ccccc}
\hline\hline
    \multicolumn{5}{c}{EM helicity amplitudes  ~[$10^{-3} \times$ GeV$^{-\frac{1}{2}}$] }                                   \\
                                   &\hspace*{.5cm}  Empirical  \hspace*{.5cm} &\hspace*{.5cm} Fit 1  \hspace*{.5cm}     &  \hspace*{.5cm} Fit 2  \hspace*{.5cm}    & \hspace*{.5cm} $\eta(B^*)$ \hspace*{.5cm}\\
\hline
         $A^p_{\frac{1}{2}}[N(1535)]$      &90(30)  &90.0   &92.8   &1  \\
         $A^n_{\frac{1}{2}}[N(1535)]$      &-46(27) &-46.0  &-47.7 &1  \\[2mm]
\hline
         $A^p_{\frac{1}{2}}[N(1520)]$      &-24(9)  &-24.0  &-26.1  &-1 \\
         $A^n_{\frac{1}{2}}[N(1520)]$      &-59(9)  &-59.0  &-57.5  &-1 \\
         $A^p_{\frac{3}{2}}[N(1520)]$      &150(15) &150  &151  &-1 \\
         $A^n_{\frac{3}{2}}[N(1520)]$      &-139(11)&-139 &-142 &-1 \\[2mm]
\hline
         $A^p_{\frac{1}{2}}[N(1650)]$      &53(16)  &53.0   &43.0   &-1 \\
         $A^n_{\frac{1}{2}}[N(1650)]$      &-15(21) &-15.0  &-17.5  &-1 \\[2mm]
\hline
         $A^p_{\frac{1}{2}}[N(1700)]$      &-18(13) &-18.0  &-17.2  &1  \\
         $A^n_{\frac{1}{2}}[N(1700)]$      &0(50)   &0.0    &30.9   &1  \\
         $A^p_{\frac{3}{2}}[N(1700)]$      &-2(24)  &-2.0   &2.5    &1  \\
         $A^n_{\frac{3}{2}}[N(1700)]$      &-3(44)  &-3.1   &14.2   &1  \\[2mm]
\hline
         $A^p_{\frac{1}{2}}[N(1675)]$      &19(8)   &19.0   &13.0   &-1 \\
         $A^n_{\frac{1}{2}}[N(1675)]$      &-43(12) &-43.0  &-40.0  &-1 \\
         $A^p_{\frac{3}{2}}[N(1675)]$      &15(9)   &15.0   &18.3   &-1 \\
         $A^n_{\frac{3}{2}}[N(1675)]$      &-58(13) &-58.0  &-56.6  &-1 \\[2mm]
\hline
         $A^N_{\frac{1}{2}}[\Delta(1620)]$ &27(11)  &27.0   &27.2   &1  \\[2mm]
\hline
         $A^N_{\frac{1}{2}}[\Delta(1700)]$ &104(15) &104  &96.7   &1  \\
         $A^N_{\frac{3}{2}}[\Delta(1700)]$ &85(22)  &85.0   &95.5   &1  \\[2mm]
\hline \hline
\end{tabular}
\caption{ Results for EM helicity amplitudes  from the fits to Set 1. The last column gives the sign of the corresponding strong amplitude needed in Eq. (\ref{ael2}). }
\label{resfitb}
\end{table}
\end{center}

\begin{center}
\begin{table}[h]
\begin{tabular}{cccc} \hline\hline ~~~ Excited proton ~~~&   ~$(S^*=\frac{1}{2},S^*=\frac{3}{2})$ content  ~~& ~~Moorhouse rule\\ \hline
 p(1535) & $(0.90,~0.43)$ &  not suppressed \\
  p(1650) & $(-0.43,~0.90)$ &  \checkmark \\
   p(1520) & $(0.99,~-0.13) $& not suppressed   \\
    p(1700) & $(0.13,~0.99) $&  \checkmark \\
     p(1675) & $(0.0,~1.0) $&  \checkmark \\
  \hline\hline
\end{tabular}
\caption{Test of the Moorhouse rule in the proton helicity amplitudes. }
\label{moor}
\end{table}
\end{center}

\begin{center}
\begin{table}[h]
\begin{tabular}{cclcc} \hline\hline ~~~ $n$-Body ~~~&   & ~~~~~~~~~Operator  & Order in $1/N_c$\\ \hline
 0     & $M_1=$ &$N_c\mathbf{I}$&                                                                    $-1$                     \\
 [2mm] \hline
1    & $M_2=$ &$\frac{6}{5}\sqrt{6}\left(ls\right)^{[0,0]}$&                                                                                         \\
[2mm]
2    & $M_3=$ &$\frac{144}{5}\sqrt{6}\frac{1}{N_c}\left(l^{(2)}\left(gG_c\right)^{[2,0]}\right)^{[0,0]}$&                         $0$                       \\
[2mm] 2    & $M_4=$
&$2\sqrt{\frac{2}{3}}\left(-\left(ls\right)^{[0,0]}+\frac{12}{N_c+3}
\left(l\left(tG_c\right)^{[1,0]}\right)^{[0,0]}\right)$&                                                         \\
[2mm] \hline
2    & $M_5=$ &$\frac{9}{5}\frac{1}{N_c}\left(lS_c\right)^{[0,0]}$&                                                                          \\
[2mm]
2    & $M_6=$ &$-\frac{9}{2\sqrt{2}}\frac{1}{N_c}\left(S_cS_c\right)^{[0,0]}$&                                                                        \\
[2mm]
2    & $M_7=$ &$3\sqrt{3}\frac{1}{N_c}\left(sS_c\right)^{[0,0]}$&                                                                          \\
[2mm]
2   & $M_8=$ &$6\sqrt{6}\frac{1}{N_c}\left(l^{(2)}\left(sS_c\right)^{[2,0]}\right)^{[0,0]}$&                                  $1$                   \\
[2mm] \hline\hline
\end{tabular}
\caption{Mass basis operators. Here the notation $l^{(2)}=\left(ll\right)^{[2,0]}$ is used.}
\label{masop}
\end{table}
\end{center}

\begin{center}
\begin{table}[h]
\begin{tabular}{ccrlcc}\hline \hline
 Meson/Partial wave~~~ & $n$-Body   &   &  Operator & Order in $1/N_c$ \\ \hline
    &  1    &   $O_1^{[0,1]}=$ &
$\left(\xi \ g\right)^{[0,1]}$ &
0 \\[2mm] \cline{2-5}
 $\pi$   &       &  $O_2^{[0,1]}=$ &
$\frac{1}{N_c} \ \left(\xi \left(s\
T_c\right)^{[1,1]}\right)^{[0,1]} $
& 1 \\[2mm]
 S-wave   &  2    &  $O_3^{[0,1]}=$ &
$\frac{1}{N_c} \ \left(\xi  \left(t\
S_c\right)^{[1,1]}\right)^{[0,1]} $
& 1 \\[2mm]
    &       &  $O_4^{[0,1]}=$ &
$\frac{1}{N_c} \ \left(\xi \left(g \
S_c\right)^{[1,1]}\right)^{[0,1]} $
& 1 \\[2mm] \hline

     &   1   & $O_1^{[2,1]}=$ &
$\left(\xi \ g \right)^{[2,1]} $
& 0    \\[2mm] \cline{2-5}
     &       & $O_2^{[2,1]}=$ &
$\frac{1}{N_c} \ \left(\xi \left(s \ T_c\right)^{[1,1]}
\right)^{[2,1]} $
& $1$ \\[2mm]
$\pi$ &       & $O_3^{[2,1]}=$ &
$\frac{1}{N_c}  \ \left(\xi \left(t \ S_c\right)^{[1,1]}
\right)^{[2,1]} $
& $1$ \\[2mm]
 D-wave    &   2   & $O_4^{[2,1]}=$ &
$\frac{1}{N_c} \ \left(\xi \left(g \ S_c\right)^{[1,1]}
\right)^{[2,1]} $
& $1$ \\[2mm]
     &     & $O_5^{[2,1]}=$ &
$\frac{1}{N_c} \ \left(\xi  \left( g \ S_c \right)^{[2,1]}
\right)^{[2,1]} $
& $1$ \\[2mm]
     &     & $O_6^{[2,1]}=$ &
$\frac{1}{N_c} \ \left(\xi  \left( s \ G_c\right)^{[2,1]}
\right)^{[2,1]} $
& $0$ \\[2mm] \cline{2-5}
     &  3    & $O_7^{[2,1]}=$ &
$ \frac{1}{N_c^2} \ \left(\xi \left( s \left( \left\{ S_c, G_c
\right\} \right)^{[2,1]} \right)^{[2,1]}\right)^{[2,1]} $
& $1$ \\[2mm]
     &     & $O_8^{[2,1]}=$ &
$\frac{1}{N_c^2} \ \left(\xi \left( s \left( \left\{ S_c, G_c
\right\} \right)^{[2,1]} \right)^{[3,1]}\right)^{[2,1]} $
& $1$ \\[2mm] \hline
$\eta$   &    1   &  $O_1^{[0,0]}=$ &
$\left( \xi \ s \right)^{[0,0]} $
& 0 \\[2mm] \cline{2-5}
S-wave  &    2   &  $O_2^{[0,0]}=$ & $\frac{1}{N_c} \left(\xi
\left(s \ S_c\right)^{[1,0]} \right)^{[0,0]} $
& 1 \\[2mm] \hline
$\eta$      &  1    & $O_1^{[2,0]}=$ &
$\left( \xi \  s \right)^{[2,0]} $
&  0 \\[2mm] \cline{2-5}
 D-wave    &  2    & $O_2^{[2,0]}=$ &
$\frac{1}{N_c} \left(\xi \left(s \ S_c\right)^{[1,0]}
\right)^{[2,0]} $
& 1  \\[2mm]
        &       & $O_3^{[2,0]}=$ &
$\frac{1}{N_c} \left(\xi \left(s\ S_c\right)^{[2,0]}
\right)^{[2,0]} $
& 1\\[2mm] \hline \hline
\end{tabular}
\caption{Strong decay basis operators.  The upper labels $^{[L,I]}$ denote
angular momentum and isospin and how these are coupled.  The vector $\xi$ provides the transition from the $O(3)$ (orbital) $\ell=1$ (excited baryon) to $\ell=0$ (ground state baryon). It is defined by its reduced matrix element: $\langle\ell=0\mid\xi\mid\ell=1\rangle=\sqrt{3}$ \cite{Scoccola:2007sn}.} \vspace*{1cm}
\label{strop}
\end{table}
\end{center}

\begin{center}
\begin{table}[b]
\begin{tabular}{clcc}
\hline \hline
   \multicolumn{2}{c}{Operator}  & Order in $1/N_c$  &~~~ $n$-Body   \\
\hline $E1^{(0)}_1=$ & $\left(  \xi\,
s\right)^{[1,0]}$
     &  0 &  1     \\
$E1^{(0)}_2=$ & $\frac{1}{N_c}  \left(  \xi\, \left( s\ S_c
\right)^{[0,0]}\right)^{[1,0]}$
     & 1 & 2  \\
$E1^{(0)}_3=$ & $\frac{1}{N_c}  \left(  \xi\, \left( s\ S_c
\right)^{[1,0]}\right)^{[1,0]}$
     &  1 &  2 \\
$E1^{(0)}_4=$ & $\frac{1}{N_c}  \left(  \xi\, \left( s\ S_c
\right)^{[2,0]}\right)^{[1,0]}$
     & 1 &  2 \\
\hline $E1^{(1)}_1=$ & $\left(  \xi\, t\right)^{[1,1]}$
     &  0  & \hspace*{.2cm} 1  \hspace*{.2cm}    \\
$E1^{(1)}_2=$ & $\left(  \xi\, g\right)^{[1,1]}$
     &  0  &   1 \\
$E1^{(1)}_3=$ & $\frac{1}{N_c}  \left(  \xi\, \left(  s\ G_c
\right)^{[2,1]}\right)^{[1,1]}$
     & 0  &  2   \\
$E1^{(1)}_4=$ & $\frac{1}{N_c}  \left(  \xi\, \left(  s\ T_c
\right)^{[1,1]}\right)^{[1,1]}$
     &  1 &  2 \\
$E1^{(1)}_5=$ & $\frac{1}{N_c}  \left(  \xi\, \left( s\ G_c
\right)^{[0,1]}\right)^{[1,1]} + \frac{1}{4\sqrt3} \ E1^{(1)}_1$
     &  1 &  2 \\
$E1^{(1)}_6=$ & $\frac{1}{N_c}  \left(  \xi\, \left(  s\ G_c
\right)^{[1,1]}\right)^{[1,1]}+ \frac{1}{2\sqrt2} \ E1^{(1)}_2$
     &  1 & 2 \\
 \hline
$M2^{(0)}_1=$ & $\left(  \xi\, s\right)^{[2,0]}$
     &  0 & 1    \\
$M2^{(0)}_2=$ & $\frac{1}{N_c}  \left(  \xi\, \left( s\ S_c
\right)^{[1,0]}\right)^{[2,0]}$
     &  1 & 2  \\
$M2^{(0)}_3=$ & $\frac{1}{N_c}  \left(  \xi\, \left( s\ S_c
\right)^{[2,0]}\right)^{[2,0]}$
     &  1 & 2  \\
\hline $M2^{(1)}_1=$ & $\left(  \xi\, g\right)^{[2,1]}$
     &  0  & 1     \\
$M2^{(1)}_2=$ & $\frac{1}{N_c}  \left(  \xi\, \left(  s\ G_c
\right)^{[2,1]}\right)^{[2,1]}$
     &  0 & 2   \\
     $M2^{(1)}_3=$ & $\frac{1}{N_c}  \left(  \xi\, \left(  s\ T_c \right)^{[1,1]}\right)^{[2,1]}$
     & 1 &  2 \\
$M2^{(1)}_4=$ & $\frac{1}{N_c}  \left(  \xi\, \left(  s\ G_c
\right)^{[1,1]}\right)^{[2,1]}+  \frac{1}{2\sqrt2} \ M2^{(1)}_1$
     & 1 &  2 \\
\hline $E3^{(0)}_1=$ & $\frac{1}{N_c}  \left(  \xi\, \left(
s\ S_c \right)^{[2,0]}\right)^{[3,0]}$
     &  1 & 2   \\
\hline $E3^{(1)}_1=$ & $\frac{1}{N_c}  \left(  \xi\, \left(
s\ G_c  \right)^{[2,1]}\right)^{[3,1]}$
     &  0 & 2  \\
\hline \hline
\end{tabular}
\caption{EM basis operators.  The upper notation $X^{(0)}$ and $X^{(1)}$ indicates the isospin of the operator. The  NLO
operators  $E1^{(1)}_5$,   $E1^{(1)}_6$, and  $M2^{(1)}_4$ involve
linear combinations with LO operators in order to eliminate their
projections onto the LO  operators.  } \vspace*{2mm}
\label{helope}
\end{table}
\end{center}

\newpage

\end{document}